\documentclass[twocolumn, prb, showpacs]{revtex4-1}

\usepackage{graphicx}
\usepackage{dcolumn}
\usepackage{bm}
\usepackage{physics}
\usepackage{verbatim}
\usepackage{amsmath}
\usepackage{mathtools}
\usepackage{placeins}
\usepackage{afterpage}
\usepackage{float}
\usepackage[usenames,dvipsnames]{color}

\definecolor{aaltoOrange}{RGB}{255,121,0}%

\newcommand{\mh}{\mathcal{H}}

\begin{document}

\title{Particle conservation in the single-particle Green's function}

\author{Marc Dvorak}
\affiliation{Department of Applied Physics, Aalto University School of Science, 00076-Aalto, Finland}
\email{marc.dvorak@aalto.fi}

\date{\today}
\begin{abstract}
We argue that the exact single-particle Green's function ($G$) in quantum many-body theory does not conserve particle number because the single-particle basis is incomplete. We conclude that the exact $G$ is not a probability amplitude and is not $\Phi$-derivable in the Kadanoff-Baym sense. This sets up a number of inconsistencies involving normalization, the definition of $G$, interpretation of the spectral function, and $\Phi$-derivability. Our result suggests that, in the most general case and in the most literal sense, $G$ is not suitable for computing particle addition/removal spectra.
\end{abstract}

\maketitle

The single-particle Green's function ($G$) in quantum many-body theory \cite{Fetter/Walecka} describes particle addition/removal to a many-body system. From $G$, one can formulate an effective, energy-dependent problem describing the propagation of an added or removed particle. Dyson's equation is the critical piece that allows one to reduce the effective degrees of freedom from $N \pm 1$ to 1 by introducing the energy-dependent proper self-energy.

The time-ordered, single-particle Green's function at zero temperature is \cite{Fetter/Walecka,interacting_electrons_reining} 
\begin{equation}
  G(1,2)= (-i) \frac{ \bra{\Psi_0}  \hat{T} \, [ \, \hat{\psi}(1) \, \hat{\psi}^\dagger(2) \, ]   \ket{\Psi_0} }{\langle \Psi_0 | \Psi_0 \rangle} \; ,   \label{def:G}
\end{equation}
where $\hat{T}$ is the time-ordering operator, $\ket{\Psi_0}$ is the interacting ground state, and $\hat{\psi}^{\dagger}(2)$ ($\hat{\psi}(1)$) is a Heisenberg creation (annihilation) field operator. Here, numbers represent points in space and time, $1 = (\mathbf{r}_1, t_1)$, and we omit spin variables. $G$ can be interpreted in terms of particle addition/removal and be linked to spectroscopic measurements of the interacting many-body system. We are primarily concerned with the adiabatic, nonrelativistic electronic Hamiltonian of condensed matter and quantum chemistry,
\begin{equation}
\mh = \sum_{ij} t_{ij} \, a_{i}^{\dagger} a_j + \frac{1}{2} \sum_{ijkl} v_{ijkl} \, a_{i}^{\dagger}  a_{j}^{\dagger} a_l a_k, \label{secondquant}
\end{equation}
for one-body matrix elements $t_{ij}$, two-body matrix elements $v_{ijkl}$, and fermionic mode creation (annihilation) operators $a_i^{\dagger}$ ($a_i$). Our arguments and results are general to any nonrelativistic interacting system, however, and can be extended beyond Eq.~\ref{secondquant}.

\begin{figure}[h]
\begin{centering}
\includegraphics[width=\columnwidth]{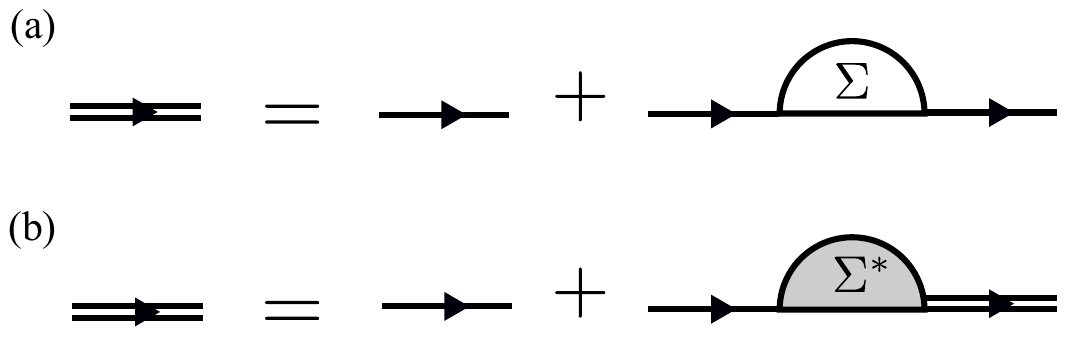}
\caption{The amplitude equation (a) and Dyson's equation (b). \label{dyson_fig}}
\end{centering}
\end{figure}

With perturbation theory, one can calculate the exact $G$ in a series based on the noninteracting single-particle Green's function, $G_0$, and matrix elements $v_{ijkl}$. The perturbation expansion for $G$ can be represented graphically in the Feynman diagram language as in Fig.~\ref{dyson_fig}a. The self-energy $\Sigma$ represents all possible contractions generated from the initial/final field operators, as prescribed by Wick's theorem \cite{wick_pr}.

Dyson's equation, shown in Fig.~\ref{dyson_fig}b, is a simplification of the amplitude equation in Fig.~\ref{dyson_fig}a. It says that the self-energy is reducible by $G_0$. One only needs to consider the proper, or irreducible, part of the self-energy, $\Sigma^*$, to compute the interacting Green's function. Dyson's equation is not part of the definition of $G$ and must be derived. Despite its success, to our knowledge, Dyson's equation has never been properly derived using only the Schr\"odinger equation and definitions. In textbooks focused on perturbation theory to calculate $G$, Dyson's equation is usually stated with only a partial derivation based on showing a few reducible diagrams. Dyson's equation can also be used to \textit{define} the proper self-energy as the connection between $G_0$ and $G$, but this approach cannot be considered rigorous since Dyson's equation is framed as a postulate.

Dyson's equation \textit{can} be derived by other methods and its structure linked to conservation laws \cite{karlsson_prb_94,baym_kadanoff_pr,baym_pr_127,interacting_electrons_reining,dyson_pr}. There are close relationships among Dyson's equation, the concept of $\Phi$-derivability, and conservation of particle number. Kadanoff and Baym showed that for a functional of $G$ labeled $\Phi[G]$ that meets certain symmetry conditions, a proper self-energy calculated as $\delta \Phi / \delta G$ satisfies a local continuity equation and therefore conserves particle number \cite{baym_kadanoff_pr,baym_pr_127}. This collection of techniques based on functional derivatives, external perturbing fields, and conservation laws is not part of the definition of $G$ or the Schr\"odinger equation. It is somewhat of an assumption that the exact $G$ (defined by the Schr\"odinger equation and Eq.~\ref{def:G}) has, or an assertion that it \textit{must} have, these same properties, including a locally conserved current.

In this work, we assess the $\Phi$-derivability concept and Dyson's equation from a different perspective that tests particle conservation using only exact quantities. We find that $G$ does not conserve particle number. This sets up several inconsistencies involving normalization, interpretation of the spectral function, and the framework of a $\Phi$-derivable set of Hedin's equations\cite{Hedin:1965}. Our overall conclusions are that the exact $G$ is not a probability amplitude and inserting the exact $G$ into a $\Phi$-derived set of Hedin's equations causes fundamental difficulties.

Before going any further, we clearly define what we mean by particle or norm conservation. We are interested in an isolated electronic system in equilibrium at zero temperature to which we can add or remove individual, quantized electrons. The system must have a definite and quantized electron number at all times since, in our model, the electrons are the fundamental degrees of freedom with infinite lifetime. We define norm or particle conservation to match an electron addition/removal experiment on the isolated system: after an initial electron addition/removal process, the total probability of annihilating the added particle/hole at any final time is $1$.

The single-particle Green's function in a discrete representation is
\begin{equation}
G_{ij}(t_1, t_2) = (-i) \frac{ \bra{\Psi_0} \hat{T} \, [ \,  a_i(t_1) \,  a_j^{\dagger}(t_2) \, ] \,  \ket{\Psi_0} }{ \langle \Psi_0 | \Psi_0 \rangle} \; .
\end{equation}
The initial state is not an eigenstate of $\mh$, in general. The repeated matrix multiplications from Taylor expanding the time evolution operator at internal times, $e^{ \mp i \mh (t_1 - t_2)}$, mix all configurations, or Slater determinants, in the $N \pm 1$ portion of Fock space. Fock space includes configurations with neutral particle-hole excitations in addition to an added particle/hole. We denote its dimension as $d_F$.

Define the states as:
\begin{eqnarray}
\ket{\gamma_j} &\equiv& e^{-i \mathcal{H}(t_1 - t_2)} \, a_j^{\dagger} e^{-i \mathcal{H} t_2 } \ket{\Psi_0}  \label{psi_j}   \\
\ket{\chi_i} &\equiv& a_i^{\dagger} e^{-i \mathcal{H} t_1} \ket{\Psi_0}  \label{chi_i}
\end{eqnarray}
where the state $\ket{\gamma_j}$ depends on the initial creation process $a_j^{\dagger}(t_2)$ (we only show this time ordering explicitly). $\ket{\gamma_j}$ is normalized because the time evolution is unitary and we assume that the initial state $a_j^{\dagger} e^{-i \mathcal{H} t_2 } \ket{\Psi_0}$ is normalized (we return to this assumption later). The action of $e^{-i \mathcal{H} t_2 }$ on $\ket{\Psi_0}$ is only a phase change. As stated above, the Schr\"odinger creation operator $a_j^{\dagger}$ acting on this state does not return an eigenstate of $\mathcal{H}$. We know nothing about the time evolved state $\ket{\gamma_j}$ except that it belongs to Fock space. Similarly, $\ket{\chi_i}$ depends on the state $i$ and is assumed to be normalized. The probability amplitude is their overlap
\begin{equation}
 G_{ij}(t_1,t_2) = (-i) \, \langle{\chi_i} | \gamma_j \rangle \; .
\end{equation}
The probability from this amplitude is 
\begin{eqnarray}
P_i^j &=& |\langle{\chi_i} | \gamma_j \rangle |^2   \\
&=& \langle \gamma_j | \chi_i \rangle \langle \chi_i | \gamma_j \rangle.
\end{eqnarray}

To test norm conservation, we choose an initial condition, set by $j$, and sum over all possible final states, each of which is set by $i$. The sum over final states is just summing down the column of $G$. The total probability of finding the particle after an initial creation $a_j^{\dagger}$ is
\begin{equation}
P^j = \sum_i \langle \gamma_j | \chi_i \rangle \langle \chi_i | \gamma_j \rangle \; .  \label{prob}
\end{equation}
For the total probability to be 1, the sum over outer products $\ket{\chi_i}\bra{\chi_i}$ must be the identity because we know that $\ket{\gamma_j}$ is normalized. Therefore, $P^j$ in Eq. \ref{prob} equals one if and only if
\begin{equation}
\sum_i \ket{\chi_i} \bra{\chi_i}  == \mathbf{I} \; .  \label{identity}
\end{equation}
We only need to count the number of $\ket{\chi_i}$ to see that Eq.~\ref{identity} is not true. Fock space is not spanned by any basis which is smaller than $d_F$ (the $N \pm 1$ portion). The number of individual field operators in $G$ that determine the number of $\ket{\chi_i}$, $d_G$, is less than $d_F$. Because $d_G < d_F$ in a many-body system, the equality in Eq.~\ref{identity} is not true. The sum in Eq.~\ref{prob} covers only a portion of Fock space, and the probability amplitude on the remaining configurations is lost. The arguments presented here are fundamental and use only exact quantities and definitions. We conclude that for nonzero time translations $\Delta t = |t_1 - t_2| > 0$, the single-particle Green's function as defined in Eq.~\ref{def:G} is not particle conserving.

How significant can the lost norm be? Imagine a system with a single-reference ground state but strongly-correlated in the $N \pm 1 $ space. After the initial creation/annihilation process, assume the time evolution pushes the system into a state which has weight only on configurations which have $\ge 1$ particle-hole pair in addition to the created particle or hole (or at least the vast majority of weight is on these configurations). This is allowed by the physics of the time evolution operator, which mixes all configurations, and Fock space.

\begin{figure}
\begin{centering}
\includegraphics[width=\columnwidth]{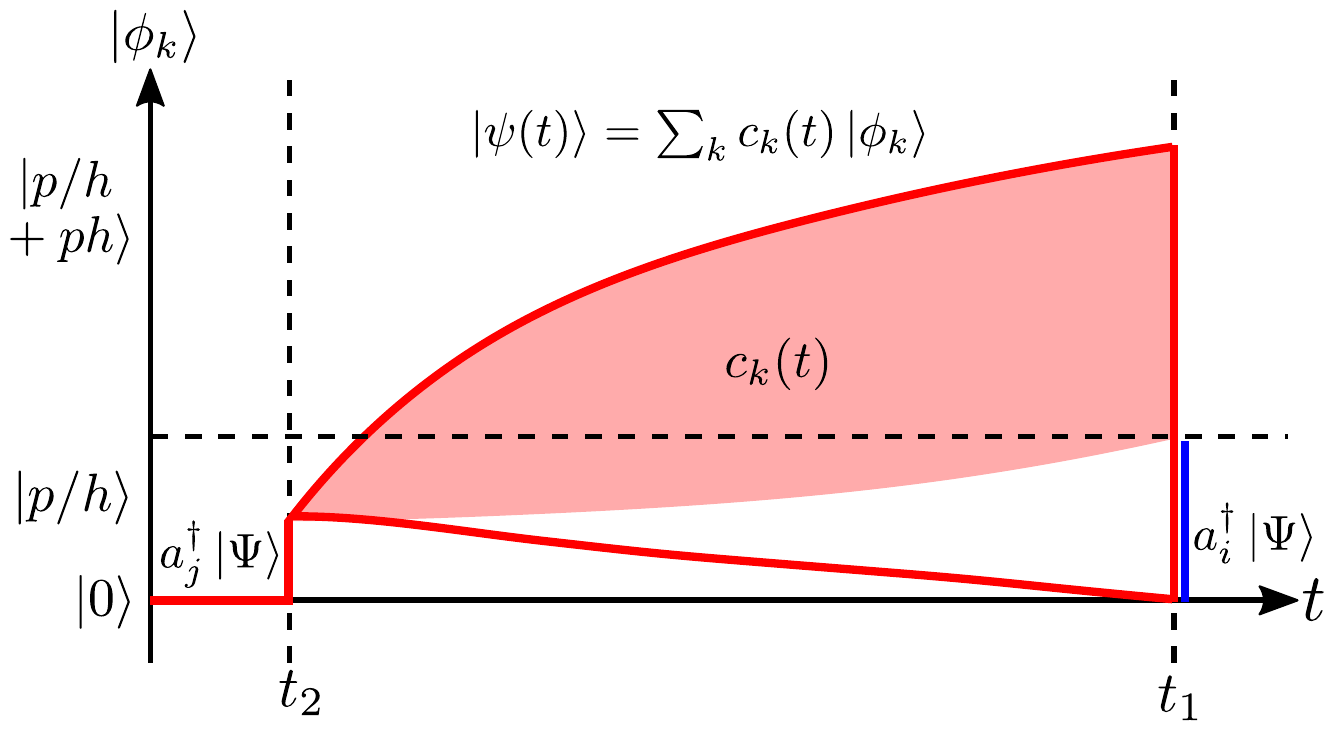}
\caption{At any time, the time dependent state $\ket{\psi(t)}$ can be expanded in the basis of configurations $\ket{\phi_k}$. The set of all $\ket{\phi_k}$ can be separated into three groups: the reference $\ket{0}$, configurations with a single particle or hole ($\ket{p/h}$), and configurations with at least one particle-hole pair ($\ket{p/h + ph}$). Assume the initial state is of the $\ket{p/h}$ type, but time evolution mixes all configurations. The nonzero expansion coefficients $c_k(t)$ for each configuration $\ket{\phi_k}$ are represented by the red outlined region. The overlap for $G$ is computed with all possible $a_i^{\dagger} \ket{\Psi}$, as indicated by the blue line. The probability amplitude above the horizontal dashed line is lost. The extreme case with zero quasiparticle residue, when the system evolves completely away from the particle/hole space, is shown by the red shaded region. \label{time_evolution}}
\end{centering}
\end{figure}

The system has evolved to the final time and we must compute the overlap with the state $a_i^{\dagger}(t_1) \ket{\Psi_0} = (\bra{\Psi_0}a_i(t_1))^{\dagger}$. However, the system has evolved into a state which cannot be connected to the ground state by a single field operator. If the ground state is the reference configuration, and the evolved state has weight only on configurations with $\ge 1$ p-h pair, the overlap between the evolved state and any state $a_i^{\dagger} (t_1) \ket{\Psi_0}$ is zero. This means that $G_{ij}(t_1,t_2)=0$ (the $j^{\mathrm{th}}$ column of $G$ is zero) at this time and the added particle is not recovered. This example is depicted graphically in Fig.~\ref{time_evolution}.

There is no guarantee that the final state can be connected to the ground state by one field operator. We do not consider the lost norm just a technical problem to be solved by reinterpreting or renormalizing the field operators in terms of quasiparticles. Arguments based on the dressing of bare particles to describe quasiparticles which are observed in experiment must be made rigorous. Eq.~\ref{secondquant} is quantized in the basis of noninteracting particles, and the field operators defining $G$ can only create or annihilate these bare degrees of freedom. Any type of normalized quasiparticle creation or annihilation, therefore, must be defined differently than Eq.~\ref{def:G}. The $G=0$ situation described here is an extreme case, but a less exaggerated time evolution and ground state is not unrealistic in strongly-correlated systems or non-Fermi liquids with zero quasiparticle residue. Such materials do exist, and proper normalization of the theory could be critical to understanding correlated phenomena.

The lost norm presents major conceptual problems. First, if $G$ is not norm conserving, it is not the correct quantity to describe particle conserving spectroscopies, even if we know the exact $G$ and regardless of any sum rule on the spectral function. The spectral function, $A(\omega) \equiv  |  \, \mathrm{Im} \, G(\omega) | / \pi $, is believed to give a probabilistic interpretation for excited states because the sum rule 
\begin{equation}
\int d\omega \, A_{ii}(\omega) = 1  \label{sum_rule}
\end{equation}
is obeyed for the exact $G$ and $A$. The sum rule, however, is different than quantum mechanical normalization. This conservation of spectral weight only states that $\langle \Psi_0 | \Psi_0 \rangle / \langle \Psi_0 | \Psi_0 \rangle= 1$ and the fermionic anticommutation relation $\{ a_i, a_j^{\dagger} \} = \delta_{ij}$. Neither of these relations imply particle conservation in $G$, and we find no proof in the definitions of $G$, $A$, or the sum rule that $G$ conserves particle number in our sense. We consider it an extrapolation from the sum rule, albeit a very successful one, that $A$ can be related to a particle conserving spectroscopy. The sum rule is only a statement about the diagonal elements of $A$. In a quantum many-body system, however, an initial state can decay into many final states. This decay is inherently an offdiagonal process, and we consider offdiagonal elements of the time evolution operator $-$ a check on all possible final states $-$ essential to proving particle conservation.

Furthermore, the initial state $a_j^{\dagger} e^{-i \mathcal{H} t_2} \ket{\Psi_0}$ meant to describe the addition of a particle is not normalized to $N+1$. Configurations contributing to the ground state can be split into two groups: those with state $j$ occupied and those with $j$ empty. The creation operator $a_j^{\dagger}$ returns 0 when acting on those configurations with occupied $j$. For these configurations, the particle number is not simply raised from $N$ to $N+1$. Instead, the amplitude is lost. If we couple to the annihilation process, $a_j$, we can make contact with those configurations that are lost in the creation process. With both $a_j^{\dagger}$ and $a_j$ processes, the initial state has some amplitude for every configuration contributing to the ground state. This is necessary for conservation of spectral weight (Eq.~\ref{sum_rule}). However, this initial state is a mixture of configurations with $N+1$ or $N-1$ electrons. It does not have a definite electron number of $N+1$ or $N-1$, as it should. This is in disagreement with our model, in which the system must have a quantized electron number at all times.

\begin{figure*}[htb]
\begin{centering}
\includegraphics[width=2.0\columnwidth]{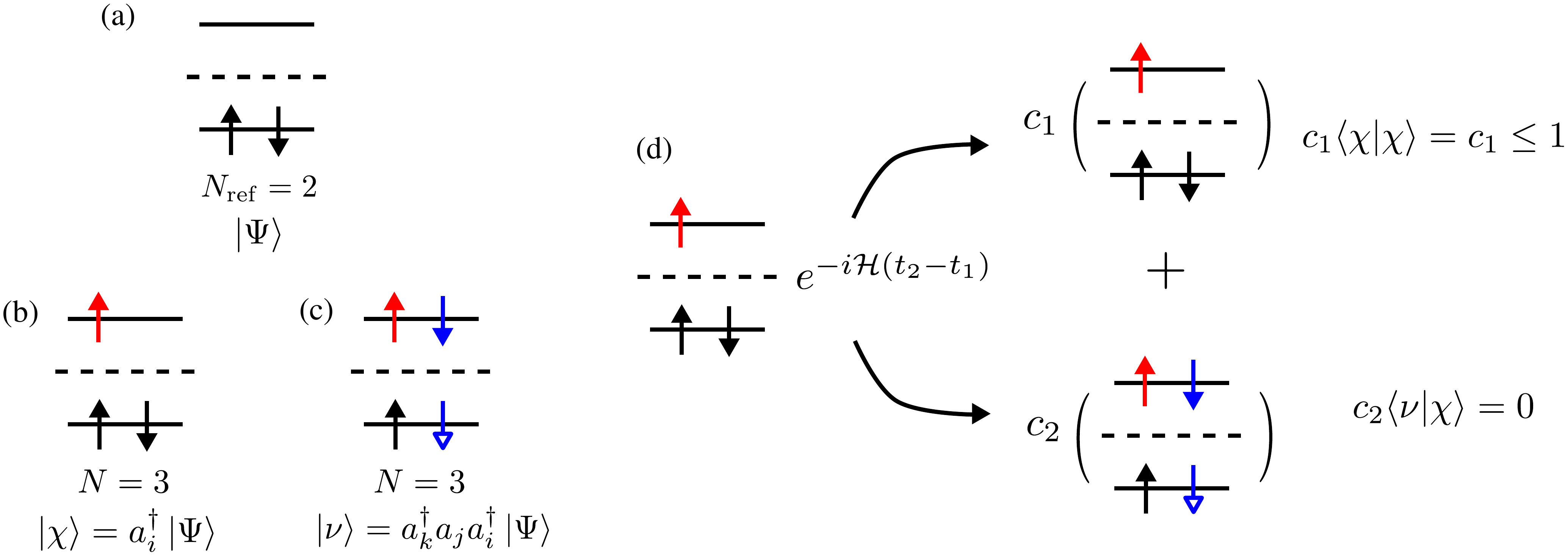}
\caption{The reference configuration (a) is a two-level system at half-filling. The added particle is shown in red (b). The virtual electron-hole pair, shown in blue, can be attached to the particle to form a trion-like state (c). The time evolution operator evolves the system into a superposition of particle and trion states (d). Only the particle amplitude, $c_1$, is recovered by the exact $G$. The trion amplitude is lost. \label{two_level}}
\end{centering}
\end{figure*}

Our overall argument has elements which are obvious and elements which are subtle, so we demonstrate the issues with a more concrete example. Consider the two-level system shown in Fig.~\ref{two_level}. We assume the ground state is a single reference configuration with two particles, shown in Fig.~\ref{two_level}a, and only consider the particle addition process. Generalizations to a multiconfigurational ground state and including the reverse time ordering do not affect our conclusions. Adding a particle to the system creates the configuration in Fig.~\ref{two_level}b labeled $\ket{\chi}$. If we forbid spin-flips, the only other configuration is shown in Fig.~\ref{two_level}c, labeled $\ket{\nu}$, and it has three particles (or rather it is defined by three field operators). $\ket{\nu}$ has a virtual electron-hole pair attached to the added particle, and we refer to it as a trion.

Because the bare particle and trion have the same particle number ($N=3$), time evolution mixes these two configurations. The time evolution from the initial state is demonstrated in Fig.~\ref{two_level}d. At the final time, the expansion coefficient for $\ket{\chi}$ is labeled $c_1$ and for $\ket{\nu}$ $c_2$. The single-particle Green's function is defined by the overlap with $\ket{\chi}$, the bare particle. The particle and trion configurations are orthogonal to each other, so only the probability from the coefficient $c_1$ is recovered.

Our argument is simple $-$ the single-particle basis is incomplete for describing an interacting many-particle system. Yet, $G$ is meant as a probability amplitude,\cite{Fetter/Walecka} which requires a normalized set of final states. The literal probability amplitude interpretation of $G$ disagrees with the incompleteness of the single-particle basis. Alternatively, one can consider the spectral function $A(\omega)$ as the fundamental object. The exact $A(\omega)$ is normalized. For our two-level system, the bare particle added to the exact ground state may overlap with a trion-like eigenstate and produce a peak in the spectral function. However, such a strong Lehmann amplitude does not guarantee that the spectral weight on the trion state can be recovered at the final time. Even our $G(t_1,t_2)=0$ example \textit{has} a normalized spectral function $-$ both properties are allowed by the definition of $G$. Evidently, the two concepts (a normalized spectral function and a normalized set of final states) are not consistent with each other. To this point, we add that normalization of the exact $A(\omega)$ is based only on definitions and not related to Dyson's equation or details of its $\Sigma^*$.

The problem is quite open to interpretation. Is such a literal interpretation of $G$ really necessary? We argue that a literal interpretation is necessary, and the current theory is incomplete. The intent of $G$ is to describe a two-point correlation and model addition or removal of normalized particles. While the spectral function certainly contains much information and has meaning as a transition rate, inspecting the spectral function does not adequately model such a process. Calculations of the many-body wave function are so successful that we consider the Schr\"odinger equation to be correct; we also know that observed particles are always normalized. To our satisfaction, there is not a theory which satisfies both properties to model two-point correlations in a way that is consistent with the Schr\"odinger equation. Our stance is very strict: we must exactly follow the Schr\"odinger equation, definition of the field operators, and definition of Fock space to describe the addition/removal of normalized particles. We are formulating a new theory with these rules and will present our results in future work.

Returning to the $\Phi$-derivability concept, the situation becomes more confusing. Based solely on the definitions, $G$ does not conserve particle number but has a normalized spectral function. If, instead, we construct a $\Phi$-derivable object that conserves particle number in the particle/hole basis, its other properties are somewhat unknown. As we have already demonstrated, the two concepts are not necessarily consistent with each other. For a $\Phi$-derivable object that we label $G^{\Phi}$, it is not clear what $|  \, \mathrm{Im} \, G^{\Phi}(\omega) | / \pi$ means since we do not think this quantity can be exactly related to the Lehmann amplitudes of the exact $G$ or $|  \, \mathrm{Im} \, G(\omega) | / \pi$. Directly requiring particle conservation in the particle/hole basis also indicates there is already some motivation for a literal enforcement of particle conservation in the theory, beyond inspecting the spectral function.

The more typical route for testing particle conservation is with a local continuity equation,
\begin{equation}
\frac{\partial}{\partial t} \, n(\mathbf{r},t) + \nabla \cdot j(\mathbf{r},t) = 0 \; , \label{continuity}
\end{equation}
for density $n(\mathbf{r},t)$ and current density $j(\mathbf{r},t)$. Setting the LHS of Eq.~\ref{continuity} equal to zero asserts that particles are conserved. As shown by Kadanoff and Baym, the continuity equation holds for a $G$ which obeys
\begin{equation}
\int d2 \, \Sigma^*(1,2)G(2,1^+) = \int d2 \, G(1,2) \Sigma^*(2,1^+)
\end{equation}
where $1^+$ is the spatial coordinate $\mathbf{r}_1$ with a time infinitesimally later than $t_1$. From here, it can be shown that
\begin{equation}
\Sigma^*(1,1') = \frac{\delta \Phi[G] }{ \delta G(1',1^{+}) }  \label{kadanoff_baym_eq}
\end{equation}
for the functional $\Phi[G]$. The result is that $G$ obeys a Dyson equation with a $\Phi$-derivable $\Sigma^*$, and the local current is conserved.

The derivation relies on the fact that the density can be written as the equal time limit of the time-ordered $G$,
\begin{equation}
n(\mathbf{r},t) = (-i) \, G(\mathbf{r}, t, \mathbf{r}, t^+ ) \; .  \label{equal_t_density}
\end{equation}
From the lesser Green's function, which we define as $G^{<}(1,2) = (i) \bra{\Psi} \psi^{\dagger}(\mathbf{r}_2,t_2) \, \psi(\mathbf{r}_1,t_1) \ket{\Psi}$, we know that
\begin{equation}
n(\mathbf{r},t) = (-i) \, G^{<}(\mathbf{r},t,\mathbf{r},t)
\end{equation}
where the two times are \textit{exactly} equal and we obtain the \textit{exact} density. For the time-ordered $G$, the limit on $t^+$ is necessary to define the time-ordering and enforce causality $-$ $G$ is not defined for equal time arguments. The density is defined at a single time, but this is exactly the case for which $G$ is \textit{not} defined. The limit $t_2 \to t^+$  is not equivalent to the equality $t_2=t$. Therefore, $G$ and $G^{<}$ must give different quantities.

For this reason, we argue that $(-i) \, G(\mathbf{r}, t, \mathbf{r}, t^+ )$ is not the density. For an infinitesimal time translation, the density must undergo an infinitesimal change as
\begin{equation}
n^+(\mathbf{r},t) = (-i) \, G(\mathbf{r}, t, \mathbf{r}, t^+ ) \neq n(\mathbf{r},t) \label{nplus}
\end{equation}
for the infinitesimally evolved density $n^+(\mathbf{r},t)$. Again, we have used only definitions to reach this conclusion. Based on the arguments discussed in this work, the infinitesimally evolved density $n^+$ may not conserve particle number. Of course, the difference between $n$ and $n^+$ is small, as it should be for a construction based on an infinitesimal, and it may not be detectable numerically. However, adding up an infinite number of these infinitesimal changes could give the finite $-$ possibly even complete $-$ loss of charge demonstrated in Eq.~\ref{prob}. Finite time evolution is built as a sum of infinitesimal time translations. The infinitesimal on $t^+$ must matter in order for the system to evolve at all. Without a way of obtaining the density from $G$, it is not possible to place $G$ in a local continuity equation and find conditions for which $G$ conserves particle number.

If $\Phi$-derivability is the same as particle conservation in our literal sense, which we believe is the case, then $G$ is not $\Phi$-derivable. Based on \textit{only} the Schr\"odinger equation, there is no \textit{ab-initio} reason for a locally conserved current in the particle/hole basis. To us, the Schr\"odinger equation actually suggests that this is not the case. Any technique which requires a conserved current, for the Ward identity or any other reason, must belong to a different formalism. We consider the requirement of local continuity in the particle/hole basis to be a redefinition of the Green's function, field operators, and/or time evolution operator that does not agree with the Schr\"odinger equation and the second quantization procedure. Roughly speaking, the redefinition describes normalized quasiparticles. For Fermi liquids, the $\Phi$-derivability construct and exact $G$ may be close to each other, but this is not generally true.

There may exist a downfolded $G$-like quantity which does conserve particle number in the particle/hole basis and more closely describes experiments. We label this unknown, particle conserving correlation function $\overline{G}$. We have demonstrated that, by definition, $G$ loses norm: $\overline{G}$ is not the single-particle Green's function. Any downfolding would define a new quantity in place of $G$. Furthermore, Dyson's equation makes no reference to any downfolding. It is a specific statement about the exact $G$, and, as such, it is a much stricter statement than simply saying that \textit{some} downfolded quantity exists which fits into Dyson's equation. At least certain methods to derive Dyson's equation do not even use downfolding.

We also know that the exact perturbation expansion starting from $G_0$ can only produce $G$, by construction. Expecting a particle conserving $\overline{G}$ from a perturbation expansion based on $G_0$ is not consistent. We do not know what $\overline{G}$ is, but it cannot come from an exact expansion based on $G_0$ and is not equivalent to Eq.~\ref{def:G}. If the perturbation expansion is truncated to an irreducible part to form Dyson's equation, the situation becomes more confusing. The piece designated $\Sigma^*$ would be from the exact expansion, but the \textit{total} series may not actually be reducible. We consider this approach to not be internally consistent.

A critical point which we cannot presently answer is whether or not our conjectured, norm conserving $\overline{G}$ \textit{can} somehow be created from $G_0$, and, if so, if it obeys a Dyson equation. $\overline{G}$ cannot obey the local time evolution operator, but it may fit in a Dyson equation. This is a topic we are actively researching. We do not know if $\overline{G}$ is the object of a $\Phi$-derivable set of Hedin's equations or if these equations describe yet another $G$-like object (Is $\overline{G}$ the same as $G^{\mathrm{\Phi}}$?).

Our perturbation theory discussion considers building $G$ (or trying to build $\overline{G}$) starting from $G_0$. Next, we assume we have the exact $G$ and discuss its behavior in a particle conserving theory. We can compute the exact $G$ from exact diagonalization and the Lehmannn representation. A self-consistent and $\Phi$-derived set of Hedin's equations is norm conserving. Inserting the exact $G$ into the framework of Hedin's equations is, again, not internally consistent because of the lost norm in $G$. There is now a mismatch between $G$ and the particle conserving working equations. If $G$ is computed from an initial $\Phi$-derived Dyson equation, $G$ is guaranteed to be norm conserving and the equations may be well-behaved. This approach would be self-consistent in Hedin's equations. If $G$ is instead computed from exact diagonalization, for example, $G$ could lose norm. Inserting this exact $G$ into Hedin's equations could cause strange behavior.

In conclusion, we have used only the definition of the single-particle Green's function, exact time evolution operator, and definition of Fock space to show that $G$ does not conserve particle number. Our discussion is thorough but not meant to be obvious or naive. We do not think the significance of this incompatibility has been appreciated. We believe that no $\Phi$-derivable $\Sigma^*$ can produce the exact $G$. There are inconsistencies between the definition of $G$, particle conservation, interpretation of the spectral function, and $\Phi$-derivability that we consider foundational issues for quantum many-body theory. We are formulating a particle conserving, $G$-like theory and will report our results in future work.

This work was supported by the Academy of Finland through grant no.~316347.

\bibliographystyle{apsrev4-1}
\bibliography{dyson}

\end{document}